\documentclass[aps,prl,superscriptaddress]{revtex4-1}
\usepackage[USenglish]{babel} 
\usepackage[T1]{fontenc}
\usepackage[ansinew]{inputenc}

\usepackage{lmodern} 

\usepackage{graphicx} 
\usepackage{amsmath}
\usepackage{amsthm}
\usepackage{amsfonts}
\usepackage{color}
\usepackage{subcaption}

\newcommand{\tmem}[1]{{\em #1\/}}

\newcommand{\tmop}[1]{\ensuremath{\operatorname{#1}}}
\newcommand{\tmtextit}[1]{{\itshape{#1}}}

\begin{document}

\title{The politics of physicists' social models}

\author{Pablo Jensen}
\email{corresponding author, pablo.jensen@ens-lyon.fr}
\affiliation{Institut Rh\^{o}nalpin des Systemes Complexes, IXXI, F-69342 Lyon, France}
\affiliation{Universite de Lyon, Laboratoire de Physique ENS Lyon and CNRS, 46 Rue d'Italie, F-69342 Lyon, France}

\begin{abstract}
I give an overview of the topic of this special issue, the "applications of (statistical) physics to social sciences at large". I discuss several examples of simple social models put forward by physicists and discuss their interest. I argue that while they may be conceptually useful to correct our intuitive models of social mechanisms, their relevance for real social systems is moot. What is more, since physicists have always needed to 'tame' the world inside laboratories to make their models relevant, I suggest that social modeling might be linked to human taming, a smashing political project.
\end{abstract}

\maketitle

\section{Introduction}

Why would physicists study social systems? Can they add anything to the
knowledge of social scientists, economists, or all of us, who practice social
systems every single day? One possible answer is given by physicist R{\'e}mi
Louf who recently earned a prestigious price for his PhD on the physics of
cities: Physics represents "a way of questioning the world and understanding
it", starting from observations to find the "simple mechanisms that govern
each phenomenon". Thanks to the avalanche of social data, the digital traces
left by everyone, physicists can now confront their simple models to social
reality and go beyond their "impressions", to found a new "science of cities [
...] that would provide a sufficiently precise image to guide political
choices". We can understand the enthusiasm of the young physicist to found a
true science of society, at once empirical and rigorous, that is,
mathematical.

Before getting too enthusiastic however, it may be useful to read a text published
nearly two centuries ago by Belgian astronomer Adolphe Qu{\'e}telet \cite{quetelet}. He
proclaimed the birth of a new science of crime: "If we want to acquire
knowledge of the general laws of [human criminal inclinations], we must gather
enough observations to ensure that everything that is not not purely
accidental is eliminated. [Thanks to this knowledge, we will have] the
possibility to improve men, by modifying their institutions and their habits". The reasoning is similar: from data to social laws, from laws to the improvement of society. And the parallel is even more striking given the title of Qu{\'e}telet's book : "Essay of social physics".

This book was part of a vast economic, political and scientific transformation
of European societies. Increasingly, strong states transformed their
territories and their inhabitants to make them governable from a center. They
counted populations and wealth to better enlist soldiers or collect taxes.
This control required the setting up of a legal and material infrastructure,
an investment similar to that of a road or rail network. Concretely, the
States generalized supervision tools that we take for granted today,
like the maps, the cadastre, the homogenization of the units of measurement or
the stabilization of the surnames.

The science of "statistics" is a direct consequence of this transformation
{\cite{desrosieres,hacking}}. At first, this word - derived from the Italian
\tmtextit{stato}, State - meant all the knowledge useful for governing a
country, and did not include mathematics. But in the 19th century, the scientific elite
invented computing tools capable of exploiting social data to help this centralized government of populations. Pierre-Simon
de Laplace, the great astronomer and mathematician, minister of Napoleon in
1799, developed different approaches to estimate the French population from
parcel data, because it was difficult - and expensive - to carry out a
comprehensive census. He assumed that the number of births per inhabitant was
more or less constant in the country, an assumption that he tested in some
thirty carefully selected regions to be representative of the whole territory.
It was then sufficient to count the number of births, which were well known
from parish registers, to obtain an estimate of the total population.

James Clerk Maxwell exported the statistical approach from social to physical
systems. In 1859 he published the founding article of a new branch of physics,
aptly called "statistical". He showed how to compute the properties of a gas
using those of its constituents, the atoms. Inspired by Qu{\'e}telet's
approach, he renounced the Newtonian approach - which dictated calculating the
trajectories of each particle - to switch to "statistical" properties, hoping
that individual unpredictability would be compensated for at the macroscopic
level. This allowed him to build the first rigorous bridge between the micro
and macro worlds, by deducing certain properties of the gas, such as
viscosity, from the statistical distribution of atomic velocities. Today,
physicists complete the circle, drawing inspiration from the well-supplied
toolbox of statistical physics to analyze social systems.

\section{Social physics today}

What are we talking about when we deal with the topic of this special issue :
"applications of (statistical) physics to social sciences at large"? A cursory
bibliometrics search of articles published in Web of Knowledge physics
journals using the term "social", "economics" or "econophysics" in either the title or the abstract leads to
roughly 9000 records. Their analysis using BiblioTools {\cite{grauwin}} reveals seven main research directions (a detailed description is given as Supplementary Information) : complex networks \cite{Boccaletti,Pastor-Satorras}, econophysics \cite{mantegna,kwapien}, opinion dynamics \cite{Baronchelli}, evolutionary games \cite{perc},  community detection \cite{newman,fortunato}, collective motion \cite{vicsek} and human dynamics \cite{holme}. Overall, the domain is steadily growing, as the number of papers has been multiplied by 10 since year 2000, reaching nowadays 800 articles per year.

Most of these articles deal with \emph{simple} models. As Castellano et al \cite{castellano} recognize in their review: "there is a striking imbalance between empirical evidence and theoretical modelization, in favor of the latter. This [...] is a rather objective reflection of a disproportion in the literature on social dynamics." The imbalance can be understood easily : simple models are attractive for physicists because they are both elegant and relevant. They capture the essential mechanisms at work in a quantitative way, stripping away unimportant details, as exemplified by the archetypical Ising model for magnetic phase transitions. This simple model extracts with surgical precision the core mechanism of phase transitions, namely the collective, avalanche-like effects provoked by particle interactions, leaving aside all the obscuring "details''. Yet, simple models are relevant for real systems, because physical systems are simplified in the laboratories \cite{mit} and thanks to the idea of universality : "statistical physics brings an important added value. In most situations, qualitative (and even some quantitative) properties of large-scale phenomena do not depend on the microscopic details of the process. Only higher level features, such as symmetries, dimensionality, or conservation laws, are relevant for the global behavior." \cite{castellano}

The basic idea behind "applications of (statistical) physics to social sciences" is also summarized very clearly in
Castellano et al. {\cite{castellano}} review : "In social phenomena, the basic constituents are not particles
but humans". Then, the "statistical physics approach to social behavior"
means trying to "understand regularities at large scale as collective effects
of the interaction among single individuals, considered as relatively simple
entities". In the "initial state", "heterogeneity dominates": "left alone,
each agent would choose a personal response to a political question, a unique
set of cultural features, his own special correspondence between objects and
words". When "interactions between social agents" are added to this initial
picture, one finds the "stunning global regularities" "denoted in social
sciences as consensus, agreement, uniformity". They add that universality gives hope that simple models will be relevant: "With this concept of universality in mind, one can approach the modelization of social systems, trying to include only the simplest and most important properties of single individuals and looking for qualitative features exhibited by models."

In this paper, I will argue that, while simple models are a good tool for physical systems, their usefulness is more limited for social systems. In short, they might be useful to improve our \emph{thinking}, invalidate intuitive models, but they do not allow us to learn much about real social systems. 

\section{A useful conceptual model}

Let's give an example of how simple social models can be useful to improve our conceptualizations of social processes
{\cite{watts,seuil}}. The segregation model proposed by
Schelling {\cite{schelling}} became one of the most studied models in social
physics, as it helps understanding why the collective state reached by agents
may be different from what each of them seeks individually.

I present here a simplified version of Schelling's model that lends itself to
an analytical solution {\cite{pnas}}. It represents the movement of a population of agents
in a "city", which consists of $Q \gg 1$ non overlapping blocks, also called
neighborhoods. Each block has the capacity to
accommodate $H$ agents. Initially, a number of agents $N = QH
\rho_0$ are distributed randomly over the blocks, leading to an average
density $\rho_0$. All agents share the same utility function $u (\rho)$ that
translates their preference for the density of the block where they are
located. The collective utility $U$ is defined as the sum of all agents'
utilities, $U = H \sum_{q = 1}^Q \rho_q u (\rho_q)$ and the average utility
$\tilde{u}$ per agent is $\tilde{u} = U / N$. The dynamics is the following:
at each time step, an agent and a free site in another block are selected at
random. The agent accepts to move to this new site only if its utility is
higher in this new location. Otherwise, it stays in its present block. Then,
another agent and another empty site are chosen at random, and the same
process is repeated until a stationary state is reached, i.e., until there are
no possible moves for any agent.

In {\cite{pnas}}, we have computed analytically the stationary states of such
a system for any utility function. They confirm previous results obtained by
numerical simulations showing that agents 'segregate' into crowded
neighborhoods of low utility. Specifically, for $\rho_0 = 0.4$, a utility given by $u(\rho) = 2 \rho$ for $\rho \leq 0.5$ and $u(\rho) = 2 (1 - \rho)$ for $\rho > 0.5$, the stationary density is given by a phase separation between
blocks that remain empty and blocks at a density $\rho = 1 / \sqrt{2}$,
leading to an average utility $\tilde{u} = 2 (1 - \rho) \simeq 0.586$.
This means that agents do not manage to reach the state of maximum average
utility ($\tilde{u} = 1$) by gathering in blocks at $\rho = 1 / 2$.

Our analytical calculations show that the surprising 'segregation' of agents
looking for half-filled neighborhoods arises because agents
\tmtextit{collectively} maximize not $U$ but an effective free energy that we
have called the \tmtextit{link} $L$. This state function allows to generalize
free energy to systems driven by {\tmem{individual}} dynamics. Its key
property is that, for any move, $\Delta L = \Delta u$. It is given
by the sum over all blocks $q$ of a potential $l_q$: $L = \sum_q l_q$, where
$l_q = \sum_{n_q = 0}^{N_q} u (n_q / H)$, with $N_q = H \rho_q$ is the total
number of agents in block $q$. In the large $H$ limit,
\begin{equation}
  \label{eq:def:link} l (\rho_q) \approx H \int_0^{\rho_q} u (\rho) 
  \hspace{0.17em} d \rho .
\end{equation}
The link may be interpreted as the cumulative of the individual marginal
utilities gained by agents as they progressively enter the blocks from a
reservoir of zero utility. Since agents move only when their individual
$\Delta u$ is positive, the stationary state is given by maximizing $L$ over
all possible densities $\{\rho_q \}$ of the blocks, from which no further
$\Delta u > 0$ can be found.

This analytical solution of Schelling's segregation model is conceptually
interesting because it allows a "clear quantitative demonstration [...] that
Adam Smith's invisible hand can badly fail at solving simple coordination
problems" {\cite{bouchaud}}. And this unwanted segregation is robust to
changes in model's ingredients: addition of noise, shape of utility
functions \ldots \cite{pancs} However, we have recently shown that it is
fragile with respect to the introduction of a vanishingly small concentration
of altruist agents {\cite{PhysRevLett.120.208301}}, a kind of "compositional
chaos".

The relevance of Schelling's model for real systems is less clear, because the reasons behind urban segregation are far more complex than those that any simple model can come up with {\cite{jasss,seuil}}. While the model shows
that one cannot logically deduce individual racism from global segregation, it says nothing about the actual urban segregation. And the idea of "universality" put forward by Castellano et al \cite{castellano} has not proved very fruitful in practice. There are some intriguing regularities in social data, such as Zipf's power law, but they are not very useful to understand social systems because they are too easy to obtain \cite{efkpowerlaw}.

\section{Finding the essential mechanisms}

To avoid the criticism of irrelevance while keeping the conceptual advantages
of simplicity, one interesting proposition is to link the models to real data,
hoping that these are produced by a single "essential" mechanism. The elegant
model of cities proposed by Louf \& Barth{\'e}l{\'e}my represents an exemplary
case of this strategy {\cite{PhysRevLett.111.198702}}. It explains the increase in the number of
urban centers - areas of high employment density - when the number of
inhabitants increases by a single "essential" mechanism: congestion. A small
city has only one center bringing together most companies and administrations,
while larger cities will have many, like the Parisian hubs La D{\'e}fense, Les
Halles and many others. To quantitatively link the population and the number
of centers, the model creates a virtual city in which there are several
potential employment centers, each offering a different salary. Each
inhabitant is a social atom preoccupied by one thing: choosing the employment center that offers the best compromise between (high) salary and (low) transportation cost. Clearly, when the city is small, the traffic
is low and all residents can go to the job center that offers the best salary:
there is therefore only one active center. But as the population grows,
traffic and congestion increase. As a result, centers offering slightly lower
salaries become active, because their proximity compensates for lost wages.
This model is attractive because it combines three advantages, which are
difficult to tie together: a mathematical link between ingredients and
consequences, a quantitative fit to data for 9000 U.S. cities and an intuitive
understanding of the phenomenon.

However, its relevance for real cities is moot. First, a rigorous mathematical
link between assumptions and consequences does not guarantee the interest of the result: the global rigor of a
chain of reasonings is that of its weakest link! And the choice of variables
or the simplifications that led to the model are more fragile. As the authors
acknowledge, the definition of an employment center is rather vague: should a
minimum number of jobs be required to declare that such a zone is a center? At
which value to set the threshold? Should two neighboring areas be considered
as one or two distinct centers? Moreover, bold assumptions are needed to build
such a simple model: residents and businesses are identical, choose their
residence at random, all firms in each center offer the same salary, there is
no public transportation... In short, using mathematics to produce
explanations by linking these elements is like trying to supply your home with
water by using a very solid pipe to a tank ...that is almost empty.

All things considered, the idea of "essential mechanisms" governing social
systems is as seductive as it is reckless. It postulates the existence of a
hierarchy among the many imaginable causes, which would allow to extract a
single one, that dominates all cases. In the city model, it is well-known that other factors lead to the creation of centers : companies want to be close to each other to facilitate the
exchange of goods or information; retail stores to attract a larger clientele
\ldots We can therefore imagine several simple models, with very different
"essential" ingredients, leading to satisfactory empirical predictions,
because the data are always noisy and do not allow to discriminate among them.

Social sciences have created tools that may be more adapted to the complexity
of social systems, where several causes have to be combined to produce an
effect. When causes can simply be added, as physicists' forces, old tools
such as multiple regressions can do the job. But in real systems, the combination is often
trickier. For example, a strike may start when either (1) a new technology is introduced and wages stagnate or when (2) the suppression of overtime is combined with outsourcing. Each of the four possibles causes in neither sufficient nor necessary to start a strike : for example, wage stagnation will not cause it if no new technology is introduced, and a strike may start even when wages are increased, through the second causal path. More complex causal tools are needed than "finding the essential mechanism" or even multiple regression {\cite{ragin}}. The point is that if one assumes from the start that there is a single mechanism at play, noisy data may confirm this idea, even when it is too simple. Respecting the
complexity of our object is essential for good science, and this may well mean giving up our fascination for elegant models.

\section{More realistic models?}

Physicists may draw comfort from the fact that more complex models, that try to include all the relevant
variables, fare no better in practice. For example, large teams of economists
build complicated models with hundreds of variables to predict economic
growth. Alas, these predictions are not much better than those of a much
simpler model: economic growth next year will be the same as this year
{\cite{growth}}. In {\cite{seuil}}, I discuss at length why complex social models
fail. One important point is the absence of conservation laws, which are a major provider of reliable equations for physical systems. For example,
climate science models take advantage of energy and momentum conservation laws
to build a kind of framework, which grants them robust predictions even for
long times and variable conditions. In chemistry, the nature of atoms is
conserved, whatever the complexity of the transformations. There is no similar
stability in social systems, hampering any credible prediction beyond the
simple "tomorrow will be as today". In other words, social science lacks a
{\tmem{{\tmem{{\tmem{dynamical}}}}}} theory, based not on simple regularities
but on the {\tmem{reproducibility of change}}, as in Newton's second law :
$\tmop{dv} / \tmop{dt} = F / m$.

\section{We are not social atoms!}

Let's examine this idea of "conservation laws" and analyze in more detail the example of "social atoms", a pervasive analogy in economists' (and physicists') models {\cite{cho}}. As summarized above by Castellano et al. \cite{castellano}, physicists' models start with isolated "simple entities", endowed with stable characteristics, and try to "understand the regularities at large scale" that arise when one adds "interactions" between them. 

\begin{figure}%
\includegraphics[width=.3\columnwidth]{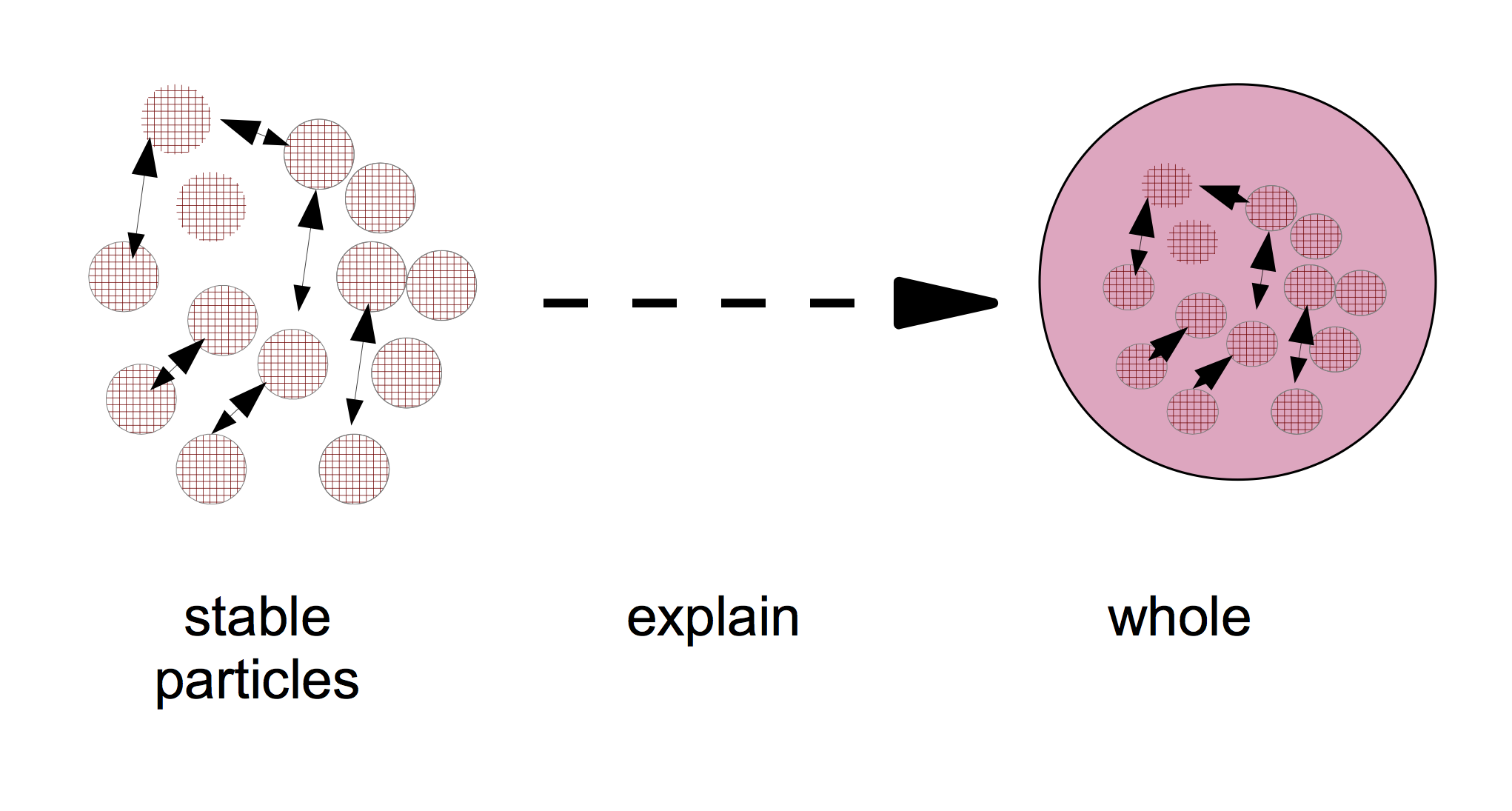}
\includegraphics[width=.3\columnwidth]{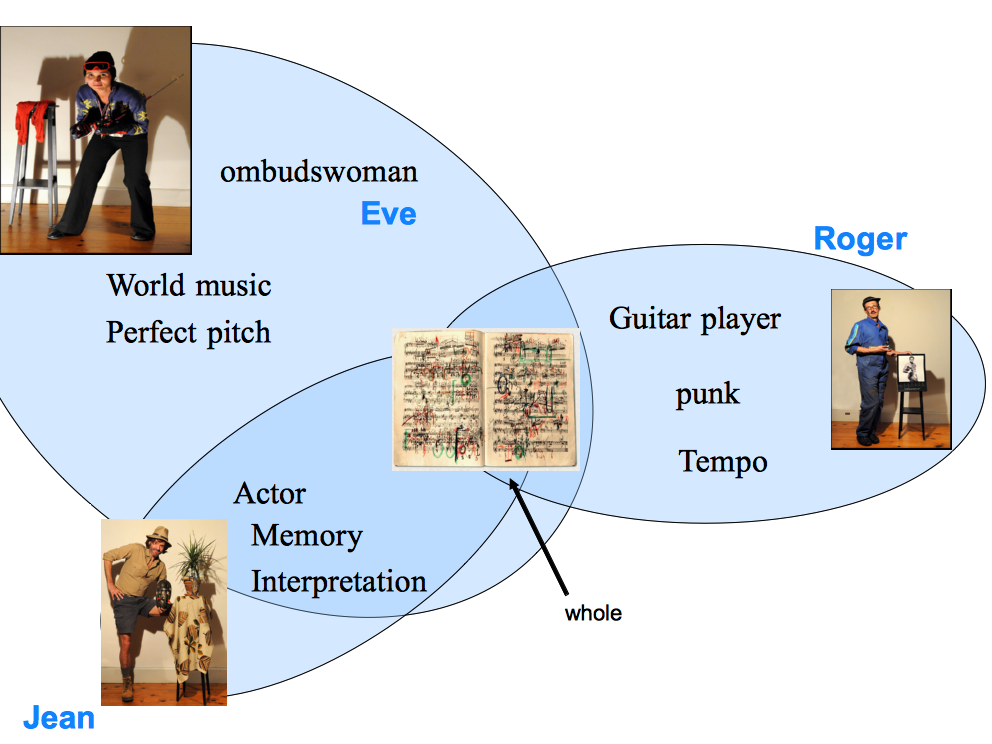}%
\caption{(Left) In the usual 'atomic' modeling, each individual is simple, stable, and their interactions 'explain' the whole, which is large as it aggregates the 'atoms'. (Right) In our vision, each individual, represented here as an ellipse with several characteristics, is complex. The whole is represented as their intersection - here, the annotated score - and is smaller than the individuals. There is no guarantee of stability in neither the individuals nor the whole. }%
\label{fig:ginga}%
\end{figure}

This approach has turned out to be fruitful to analyze physical systems, because physicists' atoms can be characterized by stable characteristics. These arise from the strong difference between the typical energy of a chemical reaction and that needed to change the chemical atomic identity, guaranteed by the much more strongly bonded atomic nucleus. The problem is that there seems to be no such energy scale separation for the so-called "social atoms", i.e. humans. Therefore, the whole idea of starting with isolated individuals endowed with stable characteristics and then adding interactions is unfounded. As argued long ago by John Dewey {\cite{dewey}}: "Each human is born an infant [...] immature, helpless, dependent upon the activities of others. There is no sense in asking how individuals come to be associated. They exist and operate in association". Moreover, human characteristics originate in these interactions : "What [a person] believes, hopes for and aims at is the outcome of association and intercourse". Unlike atoms, we are constantly remade by who and what we meet. There is no stable nucleus that would characterize us deeply, lending stability to our actions. This sheds doubts on the reliability of approaches that try to predict collective outcomes based on utility functions taken to be stable across individuals, situations or in time {\cite{seuil}}. Of course, it is always possible to build an abstract model where any macro-structure is conveniently "explained" by some arbitrarily posited stable microstructure (Fig.~\ref{fig:ginga}). Three centuries ago, Descartes' followers 'explained' the acidity of lemons by the tiny 'needles' of 'lemon atoms' \ldots

The failure of simple models does not mean that the relation between individuals and collectives
cannot be conceptualized (if not quantified). But the scope is not necessarily to stick with general models, but rather to find conceptualizations which are relevant for social systems. The interesting question becomes how humans "come to be connected in just those ways that give human communities traits so different from those which mark assemblies of electrons, unions of trees in forests, swarms of insects, herds of sheep" {\cite{dewey}}. In recent work with
sociologists {\cite{whole}}, we have discussed an alternative vision of the
entire parts/whole perspective, showing that the standard micro/macro approach
oversimplifies both the individual and global levels. For example, in
Schelling's model, individuals are defined by their color and their utility
function, which do not change during the entire process. The whole is defined
by the segregation at the (large) scale of the city, which emerges from the
interactions between individuals. However, if we cared to survey real people,
we would of course learn that each individual is more complex than her utility
function, and also that she has specific visions of both the neighborhood and
the segregation. Adolescents attending local high schools will likely have
segregation experiences that differ from adults that work abroad or retirees
who stay on the neighborhood all day. And the researcher adds her own point of
view, which is not politically neutral as we will discuss below.

We can understand these different visions of the whole with a simpler example,
that of the vocal ensemble in which I sing, Ginga, that has no leader. In the
classical vision, the parts would be the twelve singers seen as small atoms,
whose interactions would lead to the whole, the vocal group. In our approach,
each singer lends a tiny fraction of his own complexity, which he agrees to standardize, in order to create a temporary collective, a
whole much smaller than its parts. The whole is "smaller" because it obviously
represents only a small part of each of our lives. And, more important,
because the creation of a coherent group filters out the many latent
possibilities of each person. These possibilities are manyfold because we all
have a different story, in which singing takes a more or less important place.
Our musical cultures are also disparate, rather classical for some, eclectic
for others, from punk to world music. Finally, our technical skills are very
diverse, for harmony, rhythm, interpretation, pronunciation or vocal
technique. No wonder everyone has different ideas about what Ginga is, or
should be. Should we spend time recording in a studio, or focus on public concerts? Does it
really matter if we do not strictly follow the score, since few people in the public will
notice it? All these differences must, in the end, lead to a common
interpretation of each piece. A concrete trace of the "whole" would be the
annotated score (Fig.~\ref{fig:ginga}), summarizing all the musical choices
made collectively, following the more or less lively discussions that allowed
us to master the original piece. Singers are thus coordinated, partly
simplified by this standard form, the annotated score, which everyone must
respect to sing together. The whole is smaller than the parts. But it enriches
them, because none of us would have been able, individually, to produce it.

\section{Conclusion: The politics of simple models}

Let's start with a naive question : why do physicists start by stripping individual entities of their attributes, before adding simple interaction rules to obtain the "whole" that was there from the beginning? Clearly, simplifications are needed if we are to model anything. But the point is that different simplifications lead to different explanations, and in the case of social systems, to different politics {\cite{seuil,politics}}. Simple models assume that agents are unable to understand and
control collective phenomena. Implicitly, only researchers are able to analyze the situation, determine the factors leading
to the results and find the ways to change them. To use an apt image by James
Scott, the modeled individuals are, as in Taylor's factory, "the molecules of
an organism whose brain is elsewhere" {\cite{scott}}. In other words, modeling
takes an external point of view on social systems, assuming that the dynamics
of change must come from outside the situation, rather than from the
reflections and creativity of the actors {\cite{jasss,ostrom}}. This is their
implicit political vision, adapted to the control of a periphery by a center,
which needs standardized entities and relatively simple interaction mechanisms
to guide its actions. Ironically, a supposedly "bottom-up" approach \cite{epstein} leads to "top-down" social politics! Maybe it is time to, as Phil Mirowski suggests, \cite{mirowski} "abjure the physics" for the modeling of social systems. For him, the "various attempts to directly appropriate models from physics, and then bend them to the description of economic variables" have not proved successful, even if they constitute "a major source of continuity in the history of neoclassical
economics". The alternative conceptualization sketched above, inspired by sociology, keeps the complexity of individuals and their specific visions of the situation, giving
more power to the actors than to the center for analyzing and changing the collective state. 

Finally, if we analyze empirically how physics has managed to get a grip on the world \cite{pickering}, we find that it has always \emph{transformed} its objects in the laboratory, as tigers are tamed before participating in a circus show. The analogy is interesting \cite{mit}, as it shows the hard and creative work needed to adjust the theory and the world. It also stresses that there exists, at the same time, some discontinuity between nature and scientific objects (we cannot know much about the untransformed world, as we can't use a wild tiger in a circus), \emph{and} some continuity (it's a real tiger, and it will not accept to do anything). To become relevant, social models will have to 'tame' humans, for example by using the ''social credit'' system currently being developed in China \cite{china}. It aims to track people and reward ''good social behavior'' while punishing bad behavior, using monetary rewards and penalties. This may achieve better predictions, but I'm not sure that I want to contribute to the taming of humans, at least not without their consent.

\bibliography{refs}

\begin{thebibliography}{10}

\bibitem{quetelet}
A~Quetelet.
\newblock {\em Sur l'homme et le développement de ses facultés, ou Essai de
  physique sociale}.
\newblock Bachelier (Paris), 1835.

\bibitem{desrosieres}
A~Desrosieres.
\newblock {\em The Politics of Large Numbers}.
\newblock Harvard Univ Press, 2002.

\bibitem{hacking}
I~Hacking.
\newblock {\em The Taming of Chance}.
\newblock Cambridge University Press, 1990.

\bibitem{grauwin}
Sebastian Grauwin and Pablo Jensen.
\newblock Mapping scientific institutions.
\newblock {\em Scientometrics}, 89(3):943, Aug 2011.

\bibitem{Boccaletti}
S.~Boccaletti, V.~Latora, Y.~Moreno, M.~Chavez, and D.-U. Hwang.
\newblock Complex networks: Structure and dynamics.
\newblock {\em Physics Reports}, 424(4):175 -- 308, 2006.

\bibitem{Pastor-Satorras}
Romualdo Pastor-Satorras, Claudio Castellano, Piet Van~Mieghem, and Alessandro
  Vespignani.
\newblock Epidemic processes in complex networks.
\newblock {\em Rev. Mod. Phys.}, 87:925--979, Aug 2015.

\bibitem{mantegna}
R~Mantegna and HE~Stanley.
\newblock Scaling behaviour in the dynamics of an economic index.
\newblock {\em Nature}, 376:46--49, 1995.

\bibitem{kwapien}
J~Kwapien and S~Drozdz.
\newblock Physical approach to complex systems.
\newblock {\em Physics Reports}, 515(3):115 -- 226, 2012.
\newblock Physical approach to complex systems.

\bibitem{Baronchelli}
Andrea Baronchelli.
\newblock The emergence of consensus: a primer.
\newblock {\em Royal Society Open Science}, 5(2), 2018.

\bibitem{perc}
Matjaz Perc, Jillian~J. Jordan, David~G. Rand, Zhen Wang, Stefano Boccaletti,
  and Attila Szolnoki.
\newblock Statistical physics of human cooperation.
\newblock {\em Physics Reports}, 687:1 -- 51, 2017.
\newblock Statistical physics of human cooperation.

\bibitem{newman}
M.~E.~J. Newman and M.~Girvan.
\newblock Finding and evaluating community structure in networks.
\newblock {\em Phys. Rev. E}, 69:026113, Feb 2004.

\bibitem{fortunato}
Santo Fortunato and Darko Hric.
\newblock Community detection in networks: A user guide.
\newblock {\em Physics Reports}, 659:1 -- 44, 2016.
\newblock Community detection in networks: A user guide.

\bibitem{vicsek}
T~Vicsek and A~Zafeiris.
\newblock Collective motion.
\newblock {\em Physics Reports}, 517(3):71 -- 140, 2012.
\newblock Collective motion.

\bibitem{holme}
Petter Holme and Jari Saramaki.
\newblock Temporal networks.
\newblock {\em Physics Reports}, 519(3):97 -- 125, 2012.
\newblock Temporal Networks.

\bibitem{castellano}
C~Castellano, S~Fortunato, and V~Loreto.
\newblock Statistical physics of social dynamics.
\newblock {\em Rev Mod Phys}, 81:591, 2009.

\bibitem{mit}
P~Jensen.
\newblock {\em An ontology for physicists' laboratory life}.
\newblock MIT Press, 2016.

\bibitem{watts}
DJ~Watts.
\newblock {\em Everything is Obvious}.
\newblock Crown Business, 2011.

\bibitem{seuil}
P~Jensen.
\newblock {\em Pourquoi la societe ne se laisse pas mettre en equations}.
\newblock Seuil (Paris), 2018.

\bibitem{schelling}
TC~Schelling.
\newblock Dynamic models of segregation.
\newblock {\em Journal of Mathematical Sociology}, 1:143--186, 1971.

\bibitem{pnas}
S~Grauwin, E~Bertin, R~Lemoy, and P~Jensen.
\newblock Competition between collective and individual dynamics.
\newblock {\em Proceedings of the National Academy of Sciences},
  106(49):20622--20626, December 2009.

\bibitem{bouchaud}
Jean-Philippe Bouchaud.
\newblock Crises and collective socio-economic phenomena: Simple models and
  challenges.
\newblock {\em Journal of Statistical Physics}, 151(3):567--606, 2013.

\bibitem{pancs}
Romans Pancs and Nicolaas~J. Vriend.
\newblock Schelling's spatial proximity model of segregation revisited.
\newblock {\em Journal of Public Economics}, 91(1):1 -- 24, 2007.

\bibitem{PhysRevLett.120.208301}
Pablo Jensen, Thomas Matreux, Jordan Cambe, Hernan Larralde, and Eric Bertin.
\newblock Giant catalytic effect of altruists in schelling's segregation model.
\newblock {\em Phys. Rev. Lett.}, 120:208301, May 2018.

\bibitem{jasss}
T~Venturini, P~Jensen, and B~Latour.
\newblock Fill in the gap: A new alliance for social and natural sciences.
\newblock {\em JASSS}, 18(2):11, 2015.

\bibitem{efkpowerlaw}
Evelyn Fox~Keller.
\newblock Revisiting scale free networks.
\newblock {\em BioEssays}, 27(10):1060--1068, 2005.

\bibitem{PhysRevLett.111.198702}
R\'emi Louf and Marc Barthelemy.
\newblock Modeling the polycentric transition of cities.
\newblock {\em Phys. Rev. Lett.}, 111:198702, Nov 2013.

\bibitem{ragin}
Charles~C. Ragin and Beno\^{\i}t Rihoux.
\newblock {Qualitative comparative analysis (QCA): State of the art and
  prospects}.
\newblock {\em Qual. Methods}, 2(2):3--13, 2004.

\bibitem{growth}
Thomas Jobert and Lionel Persyn.
\newblock {Quelques constats sur les previsions conjoncturelles de la
  croissance francaise}.
\newblock {\em Revue d'economie politique}, 122(6):833--849, 2012.

\bibitem{cho}
Adrian Cho.
\newblock Ourselves and our interactions: The ultimate physics problem?
\newblock {\em Science}, 325(5939):406--408, 2009.

\bibitem{dewey}
J~Dewey.
\newblock {\em The Public and Its Problems: An Essay in Political Inquiry}.
\newblock Pennsylvania State University Press, 2012.

\bibitem{whole}
Bruno Latour, Pablo Jensen, Tommaso Venturini, Sébastian Grauwin, and
  Dominique Boullier.
\newblock The whole is always smaller than its parts, a digital test of gabriel
  tardes monads.
\newblock {\em The British Journal of Sociology}, 63(4):590--615, 2012.

\bibitem{politics}
B~Latour.
\newblock {\em Politics of explanation, in Knowledge and Reflexivity, ed. by
  Steve Woolgar}.
\newblock SAGE, 1988.

\bibitem{scott}
J.~C. Scott.
\newblock {\em Seeing like a state: How certain schemes to improve the human
  condition have failed}.
\newblock New Haven: Yale University Press, 1998.

\bibitem{ostrom}
E~Ostrom.
\newblock Beyond markets and states: Polycentric governance of complex economic
  systems.
\newblock {\em American Economic Review}, 100(3):641--72, June 2010.

\bibitem{epstein}
J.M. Epstein and R.~Axtell.
\newblock {\em Growing Artificial Societies. Social Science from the Bottom
  Up}.
\newblock Cambridge, Mass: MIT Press, 1996.

\bibitem{mirowski}
Mirowski P.
\newblock Markets come to bits: Evolution, computation and markomata in
  economic science.
\newblock {\em J. of Economic Behavior \& Org}, 63:209 -- 242, 2007.

\bibitem{pickering}
A~Pickering.
\newblock {\em The Mangle of Practice}.
\newblock The University of Chicago Press, 1995.

\bibitem{china}
C~Rollet.
\newblock {\em The odd reality of life under China's all-seeing credit score
  system}.
\newblock Wired, 2018.

\end{thebibliography}
\bibliographystyle{unsrt} 

\end{document}